\documentclass[seceq]{ptptex}
\usepackage{graphicx}
\usepackage{fullpage}
\notypesetlogo                       

\begin{document}
\newcommand{\beq}{\begin{equation}}
\newcommand{\beqa}{\begin{eqnarray}}
\newcommand{\eeq}{\end{equation}}
\newcommand{\eeqa}{\end{eqnarray}}
\newcommand{\non}{\nonumber}
\newcommand{\lb}{\label}
\newcommand{\fr}[1]{(\ref{#1})}
\newcommand{\E}[1]{\langle{#1}\rangle}
\newcommand{\cc}{\mbox{c.c.}}
\newcommand{\ve}{{\varepsilon}}
\newcommand{\e}{\mbox{e}}
\markboth{
S. Goto
}{
From a unstable periodic orbit to Lyapunov exponent and macroscopic variable
}

\title{From a unstable periodic orbit to Lyapunov exponent and macroscopic variable in a Hamiltonian lattice : Periodic orbit dependencies}
\author{Shin-itiro \textsc{Goto}
\inst{NTT Communication Science Laboratories, NTT Corporation, \\
2-4, Hikaridai, Seika-cho, Soraku-gun, Kyoto 619-0237, Japan}
\recdate{: Accepted for publication in Prog. Theor. Phys.}

}
%

\maketitle

\date{\today}
\begin{abstract}
We study  which and how a periodic orbit in phase space links to 
both the largest Lyapunov exponent and the expectation values of 
macroscopic variables
in a Hamiltonian system with many degrees of freedom.  
The model which we use in this paper 
is the discrete nonlinear Schr\"odinger equation.
Using a method based on  
the modulational estimate of a periodic orbit,
we predict the largest Lyapunov exponent and the expectation value of 
a macroscopic variable. 
We show that (i) the predicted largest Lyapunov exponent generally
depends on the periodic orbit which we employ, and 
(ii) the predicted expectation value of the macroscopic variable does not 
depend on the periodic orbit at least in a high energy regime. 
In addition, the physical meanings of these dependencies are considered. 

%
\end{abstract}
%

\section{INTRODUCTION}
\lb{sec:intro}

We are interested in characterizing chaotic Hamiltonian systems.    
One of the most basic indicators to characterize chaos for physicists 
is the largest Lyapunov exponent\cite{LL92}, 
and then there have been 
attempts to estimate the largest Lyapunov exponents for several systems. 
For low-dimensional Hamiltonian systems, 
Chirikov has found that the largest Lyapunov exponent is close to an averaged 
eigenvalue in a strong chaotic regime for the standard map \cite{Chi79}.
For high-dimensional Hamiltonian systems, 
an analytical method to estimate the largest 
Lyapunov exponents has been developed and applied. 
The method is based on a random approximation 
and Riemannian geometry\cite{CPC00}.
For the Hamiltonian mean field model\cite{AR95}, which 
shows a second-order transition, 
the largest Lyapunov exponent has analytically been calculated. This 
study has shown that the relation between the largest Lyapunov exponent 
and the second-order phase transition\cite{Fir98}.   
There are similar studies for the $\alpha$-XY model, which is one of the 
extended models to study how non-additivity 
affects the statistics and dynamics. 
For the $\alpha$-XY model, it has been shown 
that the largest Lyapunov exponent is a function of 
the interaction length\cite{AR98,FR01,AV02}.

It is important to estimate the expectation value of 
a macroscopic variable for a system with many degrees of freedom. 
Note here, the term  ``a macroscopic variables'' refers to a quantity obtained 
by taking the average over many degrees of freedom.
For instance, 
well-known macroscopic variables are the temperature and magnetization. 
In the equilibrium state,
the expectation values of macroscopic variables can be estimated  
using tools of equilibrium statistical mechanics 
with some basic assumptions. 
These assumptions, such as ergodicity, the principle of equal weight, 
and so on, are not usually proved. 
Then estimating the expectation value of a macroscopic 
without such classical tools sheds light on basics of statistical mechanics.

%
 
In dynamical systems theory, not only for Hamiltonian systems but for 
dissipative systems, 
utilizing and searching for periodic orbits are 
one of the most fundamental issues.
When a periodic orbit is found in a system, 
we can see a part of the skeleton of phase space 
because the periodic orbit forms an invariant subset in the phase space.
( see  Ref. \cite{PR97}, 
for the Fermi-Pasta-Ulam-$\beta$ model as an example ).

There is the question whether a periodic orbit is related to 
the largest Lyapunov exponent and the expectation values 
of macroscopic variables.  
For the Navier-Stokes equation, 
it has recently been shown that 
values of some macroscopic quantities can be estimated 
using only one unstable periodic orbit\cite{KK01,KY03,vKK06}.
In order to clarify the meanings of these numerical studies, 
it has theoretically been studied that only one unstable periodic 
orbit is enough to derive relevant statistical property 
in a hyperbolic chaotic system with many degrees of freedom\cite{KS04}.  
In addition, a method to evaluate the correlation function 
has been proposed using both one periodic 
orbit and the projection operator method\cite{KF06}.
Independently of such studies,
it has been shown that a method with only one periodic orbit can predict the 
largest Lyapunov exponent in a class 
of models including the Fermi-Pasta-Ulam-$\beta$ model\cite{DRT97}.
In Ref.\cite{DRT97}, 
they have proposed the method predicting the largest Lyapunov 
exponent in a system with $N$ degrees of freedom.
The estimate consists of the following three steps:
\begin{itemize}
\item[(1)] Find one periodic orbit $(q^{PO}(t),p^{PO}(t))$.
\item[(2)] Estimate the linear growth rate along the periodic orbit, 
$\tau_j$ ($j=1,...,2N$), and define the instability entropy 
$S_{IE}:=\sum_{j(\tau_j>0)}\tau_j$.
\item[(3)] Predict $\lambda_1=2S_{IE}/N$, based on 
the assumption $S_{IE}=S_{KS}\approx\lambda_1 N/2$, where $S_{KS}$ is the 
Kolmogorov-Sinai entropy.
\end{itemize}
On the other hand, for a macroscopic variable ${\cal O}(q,p)$, we 
could predict the expectation value in the equilibrium state. 
The method is given by the following procedure:
\begin{itemize}
\item
$\E{{\cal O}}_{PO}:=\int_0^{T_0} {\cal O}(q^{PO}(t),p^{PO}(t))dt /T_0$, 
where $T_0$ is the period of $(q^{PO}(t),p^{PO}(t))$.
\end{itemize}
This kind of substitution has also been used for 
the Fermi-Pasta-Ulam-$\beta$ model in Ref.\cite{FPC05},
where they have analytically 
shown that the largest Lyapunov exponent can also be obtained 
using an Riemannian geometric approach with this kind of substitution.

If the above two estimates depend on the 
periodic orbit that we find, 
the dependency gives a clue to understand the link
between averaged values (e.g.,the largest Lyapunov exponent and the 
expectation value of a macroscopic variable) and a 
microscopic point of view (e.g, a periodic orbit).
In general, it is difficult to discuss such a dependency,  
due to the lack of the number of periodic orbits.
As a matter of fact, in the Fermi-Pasta-Ulam-$\beta$ model, 
the number of such periodic orbits is five\cite{FPC05}.  
To discuss the dependency, we need a chaotic model in which 
many periodic orbits are easily found. If the expressions 
of these periodic orbits are analytically written, 
these expressions give analytical expressions of linear growth rates 
along the orbits.
It is noted here that there is a work in Ref.\cite{ABS06}, where
they have compared with two periodic orbits in a different point of view.  

In this paper, 
we study the periodic orbit dependencies on 
(a) the method predicting the largest Lyapunov exponent\cite{DRT97} 
consisting of the three steps mentioned above, and 
(b) substituting the expression of a periodic orbit into the definition
of a macroscopic variable. 
We treat 
the discrete nonlinear Schr\"odinger equation, in which we exactly calculate 
the modulational estimates along $N$ periodic orbits with $N$ being
the number of lattice sites. 
The fact that
there are a number of analytically expressed periodic orbits 
is a feature peculiar to the discrete nonlinear Schr\"odinger equation.
The existence of $N$ exact periodic orbits allows us to discuss the periodic orbit dependencies for (a) and (b).
Using these approach, 
we bridge some averaged quantities 
and a microscopic point of view: the largest Lyapunov 
exponent and the expectation value of a macroscopic variable  
from a periodic orbit.


\section{Theoretical Prediction and Numerical Simulation}
\lb{sec:DNLS}
The equations of motion and the Hamiltonian 
for the discrete nonlinear Schr\"odinger equation are
\beqa
\frac{du_j}{dt}&=&-i(u_{j+1}+u_{j-1}-2u_j+\gamma|u_j|^2u_j)=
\frac{\partial {\cal H}}{\partial u_j^*},\lb{eqn:EDNLS}\\
\frac{du_j^*}{dt}&=&-\frac{\partial {\cal H}}{\partial u_j},
\quad{\cal H}=i\sum_{j=0}^{N-1}(|u_{j+1}-u_j|^2-\frac{\gamma}{2}|u_j|^4),\non
\eeqa
with conditions $u_{j+N}=u_j$, $N$ being  the number of degrees of freedom.
We restrict ourselves to the case that $N$ is even. 
Here $u_j$ are  complex variables and 
$\gamma$ is a real parameter.
The system \fr{eqn:EDNLS} has the conserved quantity 
$I:=\sum_{j=0}^{N-1}|u_j|^2$,
in addition to the Hamiltonian.

In the next subsection, 
we theoretically predict the largest Lyapunov exponent and 
the expectation value of a macroscopic variable.

\subsection{Theoretical Prediction}
First, we concentrate on the Lyapunov exponent. 
we predict the largest Lyapunov exponent with the three steps
that are introduced in \S\ref{sec:intro}.

As the step (1), we find the following $N$ periodic orbits in this model, 
\beqa
u_j^{PO}(t)&=&A_k\exp\{i ( 2\pi kj/N-\omega_k(|A_k|^2)t)\},\lb{eqn:UPODNLS}\\
\omega_k(|A_k|^2)&:=&-4\sin^2(\pi k/N)+\gamma|A_k|^2,\non
\eeqa
where $k=0,...,N-1$, and $A_k\in\mathbb{C}$ are the amplitudes of the orbits. 
It is noted that 
we do not make any approximation to express these unstable periodic orbits.

As the step (2), we calculate the growth rate along $u_j^{PO}(t)$.
To calculate the growth rates of the orbits, we substitute
$u_j=u_j^{PO}(1+\mu_j),|\mu_j|\ll 1$ into Eq. \fr{eqn:EDNLS},
where $\mu_j$ are complex variables describing the tangent phase space 
along $u_j^{PO}(t)$.  
After taking into account the linear terms only and using the Fourier 
transformation,  
we obtain the linearized equations around $u_j^{PO}(t)$ in Fourier space,
\beq
\frac{d}{dt}
\left(
\begin{array}{c}
\hat{\mu}_{m-k}\\ \hat{\mu}_{k-m}^*
\end{array}
\right)
=
\left(
\begin{array}{cc}
-iB & -iC\\
iC  & iB\\
\end{array}
\right)
\left(
\begin{array}{c}
\hat{\mu}_{m-k}\\ \hat{\mu}_{k-m}^*
\end{array}
\right).\lb{eqn:MDNLS}
\eeq
Here
$$
\hat{\mu}_m:=\frac{1}{\sqrt{N}}\sum_{j=0}^{N-1}\mu_j\e^{-i2\pi mj/N},
\mu_j=\frac{1}{\sqrt{N}}\sum_{m=0}^{N-1}\hat{\mu}_m\e^{i2\pi mj/N},
$$
where $m=0,...,N-1$, and $B,C\in\mathbb{R}$ are given by 
\beqa
B&=&4\{\sin^2(\pi k/N)-\sin^2(\pi m/N)\}+\gamma|A_k|^2,\non\\
C&=&\gamma |A_k|^2.\non
\eeqa
The growth rates of the periodic orbits \fr{eqn:UPODNLS} are obtained as 
the eigenvalues of the linearized equation \fr{eqn:MDNLS}.
The squared eigenvalues are calculated as follows,
\beqa
\tau(k;m)^2&=&-16\bigg(\sin^2(\frac{\pi k}{N})-\sin^2(\frac{\pi m}{N})\bigg)^2
\non\\
&&
-8\gamma|A_k|^2\bigg(\sin^2(\frac{\pi k}{N})-\sin^2(\frac{\pi m}{N})\bigg).
\lb{eqn:LDNLS}
\eeqa
Then we have the pairs of eigenvalues $\tau_{\pm}(k;m)$.
It should be noted that the stability analysis has also been done 
without any approximation.
The expression of eigenvalues shows the following symmetries
for $\tau_\pm(k;m)$,
\beqa
&&\tau_\pm(k;k)=0,\non\\ 
&&\tau_\pm(k,N/2-m)=\tau_\pm(k,N/2+m),\non\\
&&\tau_\pm(N/2-k,m)=\tau_\pm(N/2+k,m).\lb{eqn:SymDNLS}
\eeqa
When $\tau^2(k;m)$ is positive, the periodic orbit labeled by $k$ 
is unstable. 
On the other hand, when 
$\tau^2(k;m)$ is negative, the periodic orbit is stable.
Substituting Eq. \fr{eqn:UPODNLS} into Eq. \fr{eqn:EDNLS}, we have 
the relation between the absolute value of the amplitude
and the energy density
$\epsilon:=H/N,({\cal H}=iH)$, 
$$
|A_k|^2=\sqrt{\bigg(\frac{4}{\gamma}\bigg)^2\sin^4\bigg(\frac{\pi k}{N}\bigg)
-\frac{2\epsilon}{\gamma}}
+\bigg(\frac{4}{\gamma}\bigg)\sin^2\bigg(\frac{\pi k}{N}\bigg).
$$

As the step (3), 
the linear stability analysis 
gives an analytical prediction of the largest Lyapunov exponent, 
\beq
\lambda_1=\frac{2}{N}\sum_{m=0 (\tau^2(k;m)>0)}^{N/2-1}\tau_+(k;m),
\lb{eqn:LLEDNLS}
\eeq
and this can be viewed as 
the average of the growth rates. 
Here this expression of $\lambda_1$ depends on both the parameter $k$ and
the energy density $\epsilon$. 

Second, we predict the time-averaged value of a macroscopic 
variable using a substitution of the periodic orbit. This substitution is 
introduced in \S\ref{sec:intro}.
As one of the macroscopic variables, we take
\beq
\Xi:=\frac{1}{I}\sum_{j=0}^{N-1}\frac{du_j}{dt}\frac{du_j^*}{dt}.
\lb{eqn:XiDNLS}
\eeq
Substituting the periodic orbits into Eq.\fr{eqn:XiDNLS}, 
we predict the time-averaged values of the 
macroscopic variable $\Xi$. Then we have  
$\E{\Xi}_{PO}=\omega_k^2(|A_k|^2)$.
In the high energy limit, 
we predict $\E{\Xi}_{PO}\sim\gamma^2|A_k|^4\sim(-2\epsilon\gamma)$
provided $\epsilon\gamma<0$,
and $\E{\Xi}_{PO}$ does not depend on $k$ in this limit. 
In the low energy limit, there is a periodic orbit dependency.

\subsection{Numerical Simulation and discussion}

Let us compare our predictions with numerical simulations.
We restrict ourselves to the case $\gamma<0$.
Numerical integrations of the canonical equation of motion are performed 
using a second symplectic integrator \cite{Yos98,MA92,HVA94}. 
The time step of the integrator is set at $0.005$ 
and it suppresses the relative energy error 
$\max_{0\leq t\leq 50000} [\{ (H(t)-H(0)\}/H(0) ]\sim 10^{-4}$.  
Our initial conditions of $u_j(0)$ are as follows. 
For $Re\{u_j(0)\}$, small perturbation terms are added to 
$\propto\cos(\pi j)$, and $Im\{u_j(0)\}$ are exactly zero. 
The amplitudes of $u_j(0)$ determine the value of the energy.
The computing time is set to more than $50000$ so that 
the time-averages converge, which are to obtain  
the largest Lyapunov exponent and the expectation value of 
the macroscopic variable. 
We do not take any ensemble average to obtain numerical data.

\begin{figure}
\includegraphics[width=8.5cm]
{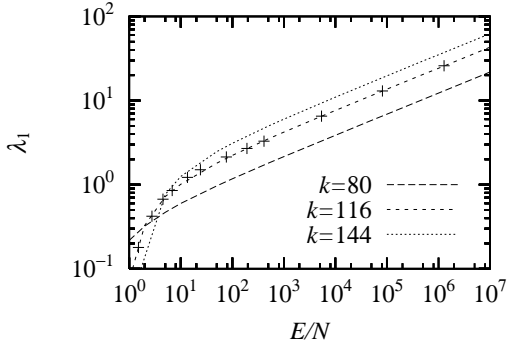}
\caption{
For 
the discrete nonlinear Schr\"odinger equation with $N=512$ and $\gamma=-1$. 
Comparison of the analytical estimate with numerical data for 
the largest Lyapunov exponents.
Pluses denote the numerically obtained the largest Lyapunov exponents and 
lines the predicted ones. 
The prediction using the periodic orbit labeled by $k=116$ 
is the best fit to the numerical data.}
\lb{fig:N512DNLS}
\end{figure}
\begin{figure}
\includegraphics[width=8.5cm]
{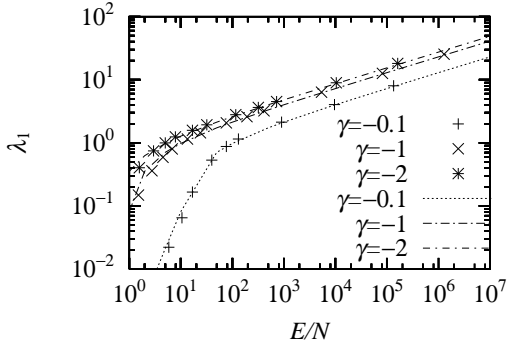}
\caption{
For the discrete nonlinear Schr\"odinger equation
with  $\gamma=-0.1,-1,-2$  and $N=128$.
Comparison of the analytical estimate with numerical data for 
the largest Lyapunov exponents.   
Pluses, crosses and asterisks denote numerically 
obtained the largest Lyapunov exponents and lines the predicted ones with 
$k=28$.
Predictions by periodic orbits labeled by $k=28$ are best fit to the numerical data.}
\lb{fig:N128gDNLS}
\end{figure}

First, we study the largest Lyapunov exponent.
In Fig. \ref{fig:N512DNLS}, we compare the largest Lyapunov exponent
$\lambda_1$ obtained both the predictions \fr{eqn:LLEDNLS} 
with the numerical calculations for the model with $N=512$. 
This figure shows that the analytical estimate depends on $k$, 
and that the prediction by taking $k=116$  is best fitted to 
the numerical data in the regime $0.1\lesssim E/N\lesssim 100$. 
We denote it by $k_*$, and 
the value of $k_*$ gives $k_*/N\sim 0.227$ for $N=512$.
For $N=128$, the best fitting parameter is $k_*=28$ (i.e., $k_*/N\sim 0.219$), 
as shown in Fig.\ref{fig:N128gDNLS}. 
Although we do not show any figure for cases $N<128$,
we find the non-trivial rule $k_*/N\sim 0.22$. 
In Fig. \ref{fig:N128gDNLS}, the $\gamma$ dependency is shown for models 
with $N=128$. 
For all $\gamma$ which we study, $k_*$ are the same each other.
Then the rule $k_*/N\sim 0.22$ can be applied 
for a wide range parameter regime.

Let us look for the origin of the non-trivial rule $k_*/N\sim 0.22$.
Here we attempt to bridge the rule for the choice of the periodic orbit 
and components of the Lyapunov vector. 
Due to the definition of the largest Lyapunov exponent, 
the origin of the rule $k_*/N\sim 0.22$ should be discussed in both
the tangent phase space and phase space, 
not only in the phase space.  
When the rule $k_*/N\sim 0.22$ is applied to the periodic orbit 
\fr{eqn:UPODNLS}, 
we can obtain the approximate periodicity in space, 
$u_{j+4,5}\sim u_j$.
We identify these values, $4\sim 5$, with the correlation length. 
On the other hand, 
it is worth to note here that the localization phenomenon 
of components of the Lyapunov vector has been studied in Ref.\cite{DRT97}.
In this paper, we calculate the normalized correlation length of 
components of the Lyapunov vector numerically as shown in Fig. \ref{fig:corrDNLS}.   
\begin{figure}
\includegraphics[width=8cm]
{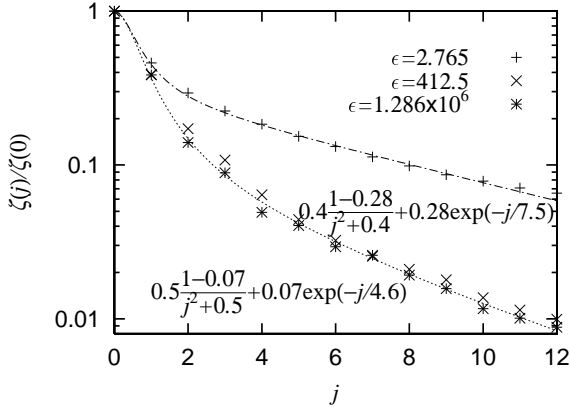}
\caption{
For the discrete nonlinear Schr\"odinger equation 
with $N=128$ and $\gamma=-1$.
Fitting the normalized correlation functions 
for the Lyapunov vector ( see Eq. \fr{eqn:CORRDNLS} )
to a sum of a Lorentzian curve and
an exponential curve.  A fitting curve gives the two 
correlation lengths, one of them is from the 
exponential curve and the other is from the Lorentzian curve. 
The correlation length is estimated as $4.6$ from the 
exponential function for the high energy regime. 
}
\lb{fig:corrDNLS}
\end{figure}
Here we define the correlation function for components of 
the Lyapunov vector as 
\beq
\zeta(j)=\lim_{t\to\infty}\frac{1}{t-t_0}\int_{t_0}^{t}dt' |du_j(t')du^*_0(t')|^2.
\lb{eqn:CORRDNLS}
\eeq
Here $t_0$ is taken as $100$ for our calculations, $du_j(t)$ denote the 
$j$-th component of the tangent vector associated with the orbit $u_j(t)$.
Fitting the normalized correlation function to 
a sum of a Lorentzian curve and an exponential curve, 
we extract the correlation length from the tangent space of the phase space. 
The fitting function which we use here is then
$$
\frac{\zeta(j)}{\zeta(0)}=\bigg(\frac{\Gamma}{2}\bigg)^2\frac{1-E}{j^2+(\Gamma/2)^2}+E\exp(-j/J),
$$
where $E,\Gamma$ and $J$ are the fitting parameters. The value of 
this function equals unity at $j=0$ for any values of $E,\Gamma$ and $J$.
Due to this fitting function, 
we have two typical length scales. 
One of them is from the width of the Lorentzian curve, $\Gamma$, and 
the other is from the exponential curve, $J$.  
The correlation lengths obtained by $J$ are approximately $4.5\sim 5$ 
for the large energy density regime.
For example, the correlation length is 4.6 
for the model with $N=128$ and $\gamma=-1$ in the high energy regime 
as shown in Fig. \ref{fig:corrDNLS}.
We can say that the correlation length in the tangent dynamics 
can be estimated using $J$. 
For the system with $N=128$ and $\gamma=-0.1$, 
although the value of $J$ is in between 
$4.5\sim 5$ in the high energy regime,
such correlation length is far from it, about $12$ in the low energy regime.
Then this explanation for the rule $k_*/N\sim 0.22$ is valid 
only in the high energy regime.
Combining the consideration of components of the Lyapunov vector and that 
of the approximate spatial periodicity by applying the rule $k/N\sim0.22$ 
to the periodic orbit,     
we could say that the emergence of the 
non-trivial rule is from the localization of the tangent dynamics 
in the high energy regime. 
It could be one of the reasons why the rule $k_*/N\sim 0.22$ appears, 
at least in the high energy regime.

Let us consider the validity of the definition of 
the instability entropy. 
If we define another version of the instability entropy, 
$S_{IE}'=\sum_{m=0~(\lambda^2(k;m)>0)}^{N-1}\lambda_+(k;m)$, 
the largest Lyapunov exponent could be predicted as 
$\lambda_1'=(2/N)\sum_{m=0 (\lambda^2(k;m)>0)}^{N-1}\lambda_+(k;m)$, 
by making the assumption that the Kolmogorov Sinai entropy is equal to $S_{IE}'$. 
This new expression $\lambda_1'$, which  is different from Eq. \fr{eqn:LLEDNLS},
cannot predict 
numerical data for the largest Lyapunov exponents
in the whole energy density regime (no figure given).
When we look for the parameter $k_*'$ 
which gives the best fit to the numerically
obtained data in the high energy regime, we find
$k_*'/N\sim 0.16$ for the models with $N=512,128$ and $\gamma=-1$.  

\begin{figure}
\includegraphics[width=8cm]
{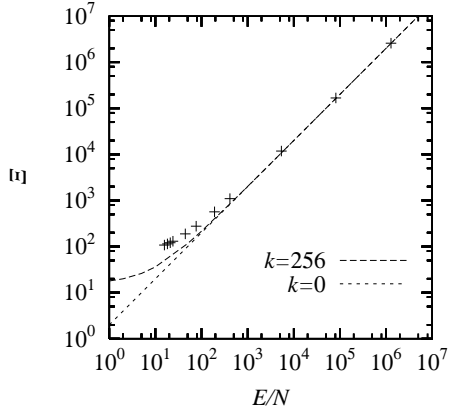}
\caption{
For the discrete nonlinear Schr\"odinger 
equation with $N=512$ and $\gamma=-1$. 
Comparison of the analytical estimate with numerical data for 
$\Xi$ ( defined in Eq. \fr{eqn:XiDNLS}).
Pluses denote the numerical data and the lines 
theoretical ones.
In the high energy regime,  the theoretical estimate by every periodic orbit   
can predict data obtained numerically.
However, in the low energy regime, the periodic orbit labeled by $k=N/2$ 
is the best fit to the numerical data, and is far from 
the numerically obtained data.}
\lb{fig:XiDNLS}
\end{figure}

Next, we compare the prediction for the time-averaged value of $\Xi$ 
with numerical simulation.
Fig. \ref{fig:XiDNLS} shows that, 
in the high enough energy regime, each prediction labeled by $k$ is in   
good agreement with numerically obtained data.
In contrast to the success of the prediction in the high energy regime,
it is difficult to conclude what happens in the low energy regime.
Although we do not show any figure, in the low energy regime 
$E/N \lesssim 100$, the time-averaged 
values of $\Xi$ do not convergent fast in time. 
However, 
from the edge of the low energy regime, $E/N \sim 50$ in Fig. \ref{fig:XiDNLS},
even the best fitted prediction, labeled by $k=N/2$, 
differs from the data obtained numerically.
Similar tendency is observed for the cases $N<512$. 

Let us look for the reason why every periodic orbit \fr{eqn:UPODNLS} 
can predict 
the time-averaged values of $\Xi$ in the high energy regime. 
Because the interaction range of this system is short, 
the system has additivity. 
For such an additive Hamiltonian system in a fully developed chaotic regime, 
we can define some macroscopic subsystems
whose sizes are macroscopically arbitrary. This kind of 
arbitrariness implies the suppression of the typical length scale, 
and then we can say that the expectation value of $\Xi$ 
does not depend on the periodic orbit characterized by $k$. 
It should be noted here that 
there is no contradiction between 
the $k$-dependency for the prediction of the largest Lyapunov exponent
and of the expectation value of the macroscopic variable. 
This is because the largest Lyapunov exponent 
is discussed in both the tangent phase space and the phase space, 
on the other hand, the expectation value of the macroscopic value is 
done only in the phase space.

We conclude that (i) the prediction of the largest Lyapunov exponent depends
on the periodic orbit, and (ii) the 
time-averaged value of the macroscopic variable can be predicted
by any unstable periodic orbit, at least in the high energy limit.

\section{CONCLUSIONS}
We have studied which and how periodic orbits predict 
both (a) the largest Lyapunov exponent and (b) the time-averaged value of
a macroscopic variable in a Hamiltonian lattice. 

To clarify these questions, 
we have studied the discrete nonlinear Schr\"odinger equation 
in which there are $N$ analytically expressed periodic orbits.
In the nonlinear Schr\"odinger equation, we have exactly constructed the 
modulational estimates along the $N$ periodic orbits and compared with
numerical simulation.
Then we have observed that the  
analytically predicted largest Lyapunov exponent depends on 
the periodic orbit and that there is a suitable periodic orbit 
for the prediction. 
The reason has been discussed 
in the  phase space and the tangent phase space by studying 
components of the Lyapunov vector.  
On the other hand, to predict the time-averaged value of a
macroscopic variable, we have substituted the analytical expression of the 
orbits into the definition of the macroscopic variable. 
The prediction is in good agreement with numerically obtained 
data in the high energy limit. 
The reason has been discussed in the phase space by considering the range of 
interactions.

In this paper, we have focused on 
a way to bridge between the dynamics and statistics.
Further investigations are necessary to quantitatively understand why 
the special periodic orbit can only predict the largest Lyapunov exponent 
even in a relatively low energy regime,  and every
periodic orbit can be used for predicting the time-averaged values of
macroscopic variables in the high energy regime. 
We believe that this kind of investigations 
can help to elucidate the study of Hamiltonian systems 
with many degrees of freedom.

\section*{Acknowledgments}
The author thanks Y.Y. Yamaguchi, M. Kawasaki, and T. Okushima,
for valuable discussions, 
S. Ruffo, for reading the manuscript and giving comments. The author also
thanks the members of the Laboratory of Dynamical Systems Theory at 
Kyoto University for their comments and encouragement, and 
the members of NTT Communication Science Laboratories for their continual
encouragement. 
The author was partially supported by a JSPS Fellowship for Young Scientists.


\begin{thebibliography}{99}
\bibitem{LL92}A.J. Lichtenberg and M.A. Lieberman, 
{\it Regular and Chaotic Dynamics} (Springer, Berlin, 1992).

\bibitem{Chi79}B.V. Chirikov,
Phys. Rep. {\bf 52}, 265 (1979).

\bibitem{CPC00}L. Casatti, M. Pettini and E.G.D. Cohen,
Phys. Rep. {\bf 337}, 237 (2000).

\bibitem{AR95}M. Antoni and S. Ruffo,
Phys. Rev. E {\bf 52} 2361 (1995).

\bibitem{Fir98}M.-C. Firpo,
Phys. Rev. E {\bf 57} 6599 (1998).

\bibitem{AR98}C. Anteneodo and C. Tsallis,
Phys. Rev. Lett. {\bf 80} 5313 (1998).

\bibitem{FR01}M.-C. Firpo and S. Ruffo,
J.Phys.A {\bf 34} L511 (2001).

\bibitem{AV02}C. Anteneodo and R.O. Vallejos,
Phys. Rev. E {\bf 65} 016210 (2002).
%
\bibitem{PR97}P. Poggi and S. Ruffo,
Physica D {\bf 103}, 251(1997).

\bibitem{KK01}G. Kawahara and S. Kida,
J. Fluid Mech. {\bf 449}, 291 (2001).
%
\bibitem{KY03}S. Kato and M. Yamada,
Phys. Rev. E {\bf 68}, 025302(R) (2003).
%
\bibitem{vKK06}L. van Veen, S. Kida and G. Kawahara,
Fluid Dyn. Res. {\bf 38}, 19 (2006). 
%
\bibitem{KS04}M. Kawasaki and S. Sasa,
Phys. Rev. E {\bf 72} 037202 (2005).

\bibitem{KF06}Miki U. Kobayashi and H. Fujisaka,
Prog. Theor. Phys., {\bf 115} 701 (2006).

\bibitem{DRT97}T. Dauxois, S. Ruffo and A. Torcini,
Phys. Rev. E {\bf 56} R6229 (1997).

\bibitem{FPC05}R. Franzosi, P. Poggi and M. Cerruti-Sola,
Phys. Rev. E {\bf 71}, 036218 (2005).

\bibitem{LPR86}R. Livi, A.Politi and S. Ruffo,
J. Phys. A {\bf 19}, 2033 (1986).

\bibitem{ABS06}C. Antonopoulos, T. Bountis and C. Skokos,
Int. J. Bifurcation and Chaos, {\bf 16}, 1777 (2006). 
%
\bibitem{Yos98}H. Yoshida,
Phys. Lett. A {\bf 150}, 262 (1998) ; Celest. Mech. Dyn. Astron. 
{\bf 56}, 27 (1998).

\bibitem{MA92}R.I. McLachlan and P. Atela,
Nonlinearity {\bf 5}, 541 (1992).


\bibitem{HVA94}B.M. Herbst, F. Varadi and M.J. Ablowitz,
Math. Comput. Simulation {\bf 37}, 353 (1994). 
%
%
%
%


%

%

\end{thebibliography}
\end{document}